\RequirePackage{fix-cm}
\documentclass[twocolumn,epjc3]{svjour3}  

\RequirePackage{graphicx}
\RequirePackage{amssymb,amsmath}
\RequirePackage{hyperref}
\RequirePackage{xcolor}

\journalname{Eur. Phys. J. C}

\begin{document}

\title{Running Einstein Constant and a Possible Vacuum State of the Universe }

\author{Giovanni Montani\thanksref{e1,addr1,addr2}
\and
Giulia Maniccia\thanksref{e2,addr2}
\and
Elisa Fazzari\thanksref{e3,addr2,addr3,addr4}
\and
Alessandro Melchiorri\thanksref{e4,addr2,addr3}
}

\thankstext{e1}{e-mail: giovanni.montani@enea.it}
\thankstext{e2}{e-mail: giulia.maniccia@uniroma1.it}
\thankstext{e3}{e-mail: elisa.fazzari@uniroma1.it}
\thankstext{e4}{e-mail: alessandro.melchiorri@uniroma1.it}

\institute{
Nuclear Department, ENEA, C.R. Frascati, Via E. Fermi 45, Frascati, 00044, Italy \label{addr1}
\and
Physics Department, Sapienza University of Rome, P.le A. Moro 5, Roma, 00185, Italy \label{addr2}
\and
Istituto Nazionale di Fisica Nucleare (INFN), Sezione di Roma, P.le A. Moro 5, Roma, 00185, Italy \label{addr3}
\and 
Physics Department, Tor Vergata University of Rome, Via della Ricerca Scientifica 1, Roma, 00133, Italy \label{addr4}
}

\date{Received: date / Revised version: date}

\maketitle

\begin{abstract}
We propose a revised formulation of General Relativity for cosmological settings, in which the Einstein constant varies with the energy density of the Universe.
We demonstrate that this modification has only phenomenological impact of providing an effective dark energy density expression. 
Assuming a state close to vacuum, here defined by the vanishing product of the Einstein coupling constant and the Universe's energy density, we perform a Taylor expansion of the theory and hence extend it to the whole domain.
In this framework, the (renormalized) vacuum energy problem is studied, and an additional constant pressure term, which induces a Chaplygin-like contribution to the dark energy sector, arises in the late-time dynamics.
The correction to the late-time Hubble parameter is investigated by comparing theoretical predictions with the late Universe observational data.
Our findings indicate that the current value of the stated vacuum energy is consistent with zero within 1$\sigma$. 
Implications of the modified $\Lambda$CDM model with respect to the Hubble tension are also discussed. 
\end{abstract}

\section{Introduction}
A characteristic feature of General Relativity (GR) is the sensitivity of the gravitational field, i.e. the space-time curvature, to the energy-momentum of any physical field. It is commonly said that gravity is an \emph{environmental} interaction \cite{Primordial,Gravitation}.
As a consequence, the gravitational field responds to any non-zero energy density, making the most natural definition of ``vacuum'' in GR the one associated to a vanishing energy-momentum tensor for all local or cosmological ``matter'' components. 

In a Minkowski space-time, the diverging energy density of a free field is renormalized to zero by prescribing that the annihilation operators appear on the right-hand side of any expression \cite{bib:mandl-shaw-qft,bib:weinberg-qft}. 
This renormalization process is partially preserved in Quantum Field Theory (QFT) on curved (or curvilinear) space-time \cite{bib:birrel-davies}, where \textit{ad hoc} renormalization schemes are developed for specific cases \cite{bib:wald-review,bib:book-wald-QFT-curved-space}. 
However, as one approaches Planck-scale physics -- i.e., exploring spatial scales of the order of the Planck length $l_P  \simeq 10^{-33} cm$ -- the renormalization process becomes inapplicable due to the emergence of quantum gravity effects (for discussions on possible quantum gravity corrections to QFT, see \cite{bib:kiefer-1991,bib:bertoni-venturi-1996,bib:casadio-2008,bib:venturi-2013-spettro,bib:kieferkramer-2013-spettro,bib:venturi-2014-spettro-slowroll,bib:brizuela-kiefer-2016-desitter,bib:brizuela-kiefer-2016-slow-roll,bib:montani-digioia-maniccia-2021,bib:kramer-chataignier-2021,bib:maniccia-montani-2022,bib:maniccia-montani-torcellini-2023,bib:maniccia-montani-antonini-2023,bib:maniccia-montani-tosoni-2024}). 

An estimate of the vacuum energy density of a free field can be calculated by assuming a cut-off value for the particle momentum of the order $1/l_P$ (in $c=\hbar =1$ units), see \cite{CQG}. 
A straightforward calculation yields a vacuum energy density of the Planck order $\propto l_P^{-4}$. 
This contribution would correspond to a massive effective cosmological constant $\Lambda$, approximately $10^{120}$ times greater than the currently estimated value \cite{bib:weinberg-grav-cosm}.

For discussions on the possible evolving nature of the present cosmological constant, see \cite{desi,giarè2024dynamicaldarkenergyplanck,Giare_Robust_DDE,montani-nakia-dainotti,bib:didonato-2025}.

Over the past 25 years, numerous studies have been conducted \cite{bib:weinberg-2000,bib:zlatev-1999,bib:mukhanov-2000,bib:padma-2002,bib:peebles-2002,bib:odintsov-2006} to explain how the immense vacuum energy is reduced to its currently observed value -- approximately $70\%$ of the present Universe's critical density. 
Within the framework of GR and the Standard Model of particle physics, this issue remains an unsolved and challenging question.
However, the picture changes significantly when modified gravity theories are taken into account, particularly the unimodular formulation (see \cite{bib:henneaux-teitelboim-1989,bib:unruh-1989,bib:kuchar-1991} for early developments and \cite{bib:percacci-2018} for a comparison between unimodular gravity and standard GR at the one-loop level).
This framework allows the gravitational field to remain unaffected by constant energy densities, by permitting $\Lambda$ to vary off-shell, with the resulting conjugate variable -- a four-volume -- being interpreted as cosmological time.

Actually, the concept of varying constants can be traced back to an initial proposal in \cite{bib:dirac-1938-varyingconst} and extends beyond just the $\Lambda$ contribution.
For instance, in \cite{bib:magueijo-2021,bib:magueijo-2022,bib:magueijo-2023}, several varying coupling constants give rise to conjugate variables that act as effective cosmological clocks.

An important outcome of the varying constants approach is to provide a potential solution to both the cosmological constant problem \cite{bib:smolin-2009} and the Hubble tension \cite{bib:divalentino-2021,bib:vagnozzi-2022,bib:vagnozzi-2020,bib:zhumabek-2024,montani_hubbletension_7,montani_hubbletension_8,montani_hubbletension_9,montani_hubbletension_10,montani_hubbletension_11}.  
In particular, one may consider an evolving vacuum energy density, that is the so-called running vacuum model \cite{bib:peracaula-2025,bib:peracaula-2023,Sol_Peracaula_2021,Dahmani2025}; see \cite{bib:peracaula-2022} for a review.
Beyond the varying constant paradigm, numerous alternative approaches have also been suggested to mitigate the Hubble tension, including different dark energy models \cite{bib:tsilioukas-2025,bib:mukhopadhyay-2025} and the backreaction of super-Hubble perturbations \cite{bib:branderberger-2025}.
For studies of combined early and late Universe modified paradigms, see \cite{bib:vagnozzi-2023}.

In this work, we consider a theory where the Einstein constant (\emph{de facto} the Newtonian constant) depends on time in a cosmological context.
Specifically, we reformulate this time dependence as a dependence of the Einstein constant on the Universe's energy density, without loss of generality.
Such function is introduced phenomenologically, in order to explore in a controlled and analytically tractable way how a variable gravitational coupling could manifest in cosmological observables.
We analyze the implications of the Bianchi identities within this revised scenario, exploring how the new physics affects both the Friedmann equations and geodesic particle motion.
We find that the $00$-component of the field equations and the $0$-component of the geodesic equations receive no additional contribution but a non-standard pressure term, due to the non-zero vacuum energy.

To explore the potential phenomenological implications of our proposal, we introduce a new definition of the \emph{macroscopic vacuum} state of the Universe: 
it corresponds to the vanishing product of the Einstein coupling constant (which depends on the energy density) and the Universe's energy density.
Assuming that the current Universe is sufficiently close to this vacuum state, we then perform a Taylor expansion of this product, leading to a specific form for the Einstein coupling.
We then extend such expansion to the whole domain of energy density values, because this way such vacuum state corresponds to a zero gravitational coupling constant.
An immediate significant phenomenological consequence arises in this limit: a constant additional pressure term appears in the present Universe's dynamics. 
When interpreted as a modification of the dark energy equation of state, this term alters the Friedmann equation; as a result, an additional contribution to the Hubble parameter appears in the form of a logarithmic term.
To quantify this effect, we employ datasets from the late Universe and use a Bayesian inference procedure to estimate the amplitude of the logarithmic term, which corresponds to twice the ratio of the vacuum energy density of the Universe to its present critical density.

The manuscript is structured as follows: in Section \ref{sec2}, we present the basic equations for perfect fluids in a curved spacetime with the varying Einstein constant. The associated Friedmann equations and Hubble parameter are derived in Section \ref{sec3}. Section \ref{sec4} introduces the vacuum definition and discusses the related energy problem. In Section \ref{sec5}, we analyze the late Universe dynamics within this framework, and compare it with observational data in Section \ref{sec6}. Finally, we discuss the results and provide concluding remarks in Section \ref{sec7}.

\section{Basic formulation}\label{sec2}

We now elucidate the fundamental paradigm underlying our proposed solution to the Hubble tension.

We consider modified Einstein equations in the presence of a space-time dependent coupling constant between gravity and matter \cite{bib:dirac-1938-varyingconst}, i.e.:
\begin{equation}
	G_{\mu\nu} = \chi (x^{\alpha})T_{\mu\nu}\, , 
	\label{htx1}
\end{equation}
where $G_{\mu\nu}$ is the usual Einstein tensor, constructed with the metric $g_{\mu\nu}$ (here we adopt the signature $(+,-,-,-)$), $T_{\mu\nu}$ denotes the matter energy-momentum tensor, and $\chi$ represents the Einstein ``constant'', here promoted to a function that depends on the event $x^{\alpha}$ 
($\mu ,\nu ,\alpha = 0,1,2,3$). 

A fundamental requirement at the ground of our phenomenological framework is the validity of the Bianchi identities, i.e. the necessity of preserving the divergenceless nature of the Einstein tensor.
The central idea we aim to implement involves a phenomenological rescaling of the gravity–matter coupling with time (or, equivalently, with the Universe’s energy density or temperature, see below), without altering the kinematical structure of the gravitational field -- that is, its geometric character as captured by second-order field equations \cite{bib:lovelock-1971}.
Accordingly, we construct the matter field dynamics by imposing that the product of the energy-momentum tensor and the running Einstein coupling constant has vanishing covariant divergence as a whole. 
This approach ensures that no violation of the foundational principles of GR occurs. 
However, the matter equations of motion may, in principle, differ from their standard form in GR, depending on the specific behavior adopted for the running coupling constant -- considered here as a free and "passive" degree of freedom within our proposal.

Thus, the validity of the Bianchi identities for $G_{\mu\nu}$ of the form \eqref{htx1}implies the following modified conservation law for the energy-momentum tensor:
\begin{equation}
	\nabla_{\nu}T_{\mu}^{\nu} = - T_{\mu}^{\nu}\partial_{\nu}\ln \chi\, , 
	\label{htx2}
\end{equation}
where $\nabla$ denotes the metric covariant derivative, and $\chi$ appears as a necessarily scalar function.

In view of the cosmological implementation of our theory, we focus on the form that Eq. \eqref{htx2} takes when matter is described by the energy-momentum tensor of a perfect fluid, which is given by \cite{bib:landau6,Primordial}
\begin{equation}
	T_{\mu\nu} = \left( \rho + p\right) u_{\mu}u_{\nu} - pg_{\mu\nu}\, , 
	\label{htx3}
\end{equation}
where $\rho$ is the energy density, $p$ is the pressure, and $u_{\mu}$ is the four-velocity field associated with matter.

Substituting Eq. \eqref{htx3} into Eq. \eqref{htx2}, and performing straightforward algebra, we obtain:
\begin{equation}
\begin{split}
	&u_{\mu}\nabla_{\nu}\left[ \left( \rho + p\right) u^{\nu}\right] + \left(\rho + p\right) u^{\nu}\nabla_{\nu}u_{\mu} \\
    &= \partial_{\mu}p - u_{\mu}\left( \rho + p\right)u^{\nu}\partial_{\nu}\chi + p \partial_{\nu}\chi \, . 
    \end{split}
	\label{htx4}
\end{equation}
Multiplying both sides of the equation by $u^{\mu}$ (with $u^{\mu}u_{\mu} = 1$), we arrive at the scalar equation
\begin{equation}
	\nabla_{\nu}\left[ \left( \rho + p \right) u^{\nu}\right] = u^{\nu}\partial_{\nu}p 
	- \rho u^{\nu}\partial_{\nu}
	\ln \chi \, ,
	\label{htx5}
\end{equation}
which, when substituted back into Eq. \eqref{htx4}, gives the equation for the fluid trajectories:
\begin{equation}
\begin{split}
	\left( \rho + p \right) u^{\nu}\nabla_{\nu}u_{\mu} =& 
	\partial_{\mu}p - u_{\mu}u^{\nu}\partial_{\nu}p \\
    &+ 
	p\left( \partial_{\mu}\ln \chi - u_{\mu} u^{\nu}\partial_{\nu}\ln \chi \right)\, .
	\label{htx6}
    \end{split}
\end{equation}

Equations \eqref{htx1}, \eqref{htx5} and \eqref{htx6} provide the basic dynamical ingredients for the cosmological analysis developed below.

\section{Cosmological Dynamics}\label{sec3}

According to the Plank Satellite data \cite{Planck2018,efstathiou_planck} we consider an isotropic flat Universe (for a different result see \cite{divalentino_planck}), with line element
\begin{equation}
	ds^2 = dt^2 - a^2(t) \delta_{ij}dx^idx^j \quad , \quad i,j=1,2,3 \, ,
	\label{htx7}
\end{equation}
where $t$ denotes the synchronous time 
(we choose $c=1$ units), $a(t)$ is the cosmic scale factor which accounts for the expansion of the Universe, and $\delta_{ij}$ is the Euclidean three-metric tensor.

Since the Universe is spatially homogeneous, the three quantities $\chi$, $\rho$ and $p$ are function of time only. We also assume the standard equation of state for the perfect fluid $p=w\rho$, where $w$ is a constant. 
Thus, the Friedmann and acceleration equations take the form, according to Eq. \eqref{htx1}:
\begin{equation}
	H^2 \equiv \left( \frac{\dot{a}}{a}\right)^2 = \frac{\chi(t)}{3}
	\sum_w \rho _w(t) \, 
	\label{htx8}
\end{equation}
and
\begin{equation}
	\frac{\ddot{a}}{a} = - \frac{\chi (t)}{6} \sum_w \left( 1 + 3w\right) \rho _w(t) \, , 
	\label{htx9}
\end{equation}
respectively. 
In these equations, the dot denotes the derivative with respect to $t$, and we allow for the presence of a generic set of energy density contributions, each characterized by a different $w$.

Now Eq. \eqref{htx5} easily provides
\begin{equation}
	\sum_w\left[ \dot{\rho}_w + 3\frac{\dot{a}}{a}\left( 1 + w \right) \rho_w \right]  = - \dot{\ln}\, \chi(t)\sum_w\rho_w\, ,
	\label{htx10}
\end{equation}
which admits the solution
\begin{equation}
	\chi(t) \sum_w\rho_w (t) = \chi_E
	\sum_w \frac{\rho_w^0}{a^{3(1+w)}}\, , 
	\label{htx11}
\end{equation}
where $\chi_E$ is the ordinary Einstein constant and $\rho_w^0$ is the present-day value of the corresponding energy density $\rho_w$  (with the convention that today $a(t=t_0) = 1$). 
Thus, the Friedmann equation \eqref{htx8} for a $\Lambda$CDM-scenario (i.e. in the presence of cold dark matter and baryonic matter
$\rho_m = \rho_m^0/a^3$ and a constant dark energy density $\rho_{\Lambda}$) can be written in the following standard form:
\begin{equation}
	H^2 (z) = H_0^2\left( \Omega^0_m(1+z)^3 + \Omega_{\Lambda}\right)
	\, , 
	\label{htx12}
\end{equation}
where $H_0\equiv H(z=0)$, $\Omega^0_m \equiv \chi_E\rho_m^0/3H_0^2$ and $\Omega_{\Lambda} \equiv \chi_E \rho_{\Lambda}/3H_0^2$. 

It is worth emphasizing that, within the proposed scenario, both the free particles motion and the Friedmann equation governing cosmological dynamics appear unchanged compared to the standard case. 
However, a more detailed analysis presented in Sec. \ref{sec5} will reveal that an additional pressure term modifies the Universe evolution. 
This unexpected contribution arises from the presence of a non-zero vacuum energy density -- indeed, the Universe asymptotically approaches a finite energy density -- and represents a key signature of our formulation. 
Notably, this new term offers the potential for empirical validation through comparison with observational data from low-redshift sources (see Sec. \ref{sec6}).

\section{Solution of the vacuum energy problem}\label{sec4}

A long standing question in GR and Cosmology \cite{weinberg_cosmologicalconstant,padmanabhan_cosmologicalconstant} concerns the possibility of defining an absolute zero for the energy density of a quantum matter field, since the Einstein equations are sensitive to any physical source including constant energy density terms. 
The present value of the cosmological constant, associated with the $\Lambda$CDM model, is extremely smaller (by about $120$ orders of magnitude) than the Planckian cut-off, which is considered the natural value for a quantum field's vacuum energy \cite{bib:weinberg-grav-cosm}.
No clear mechanisms are known, even in very general formulations \cite{sakai_vacuum_string}, to explain such a drastic suppression. The idea of a possible renormalization process \cite{bib:birrel-davies,bib:mandl-shaw-qft} is considered ambiguously applicable because, at the Planckian scale, quantum gravity effects are expected to be relevant \cite{bib:rovelli-book-QG}. 
Here we do not address this point, but we reformulate the gravitational theory in a more physical manner with respect to the role of vacuum energy.

In the present scenario, the vacuum energy process can be reformulated in a more phenomenological manner. Without loss of generality, we can choose in cosmology $\chi = \chi (\rho )$, i.e. fixing the scaling of the gravitational interaction in terms of the Universe's energy density. 
Hence, it is natural to define the \emph{a priori} vacuum state of the Universe as that one for which the following condition holds:
\begin{equation}
	\chi (\rho_{vac})\rho_{vac} = 0\, , 
	\label{htx1x}
\end{equation}
which states that the generic vacuum energy density $\rho_{vac}$ does not produce any gravitational effect on the Universe. 
A theory containing a varying Einstein constant, treated as a dynamical field, should obey a variational principle, see \cite{bib:ellis-2005}.
Here, the Einstein constant is not dynamical and thus the condition \eqref{htx1x} does not follow from an action principle. 
Such coupling resembles the running behavior of fundamental coupling constants in quantum field theory, and it is therefore formulated as an assigned phenomenological function. 
This hypothesis reflects a possibility worth exploring in light of the cosmological constant problem.

Clearly, the validity of the condition \eqref{htx1x} has a precise physical meaning if the value of the $\rho_{vac}$ is a minimum one for the Universe evolution.
Here $\rho_{vac}$ is intended as the renormalized value of a Planckian vacuum contribution, and we simply state that its presence should not influence the Universe's evolution.

Any further development of our model would require the knowledge of the explicit expression of the function $\chi (\rho )$. 
This information is not contained in the model itself, but, in what follows, we will study the relevant case in which we are close to the vacuum energy density: its value is slightly smaller than the present-day Universe critical density.
This situation has to be naturally reached by the Universe, unless the dark energy equation is exactly $p_{de}=-\rho_{de}$ (with self-explanatory notation). 
Thus, we are simply stating that our Universe has an energy density approaching the present-day vacuum value and hence, we can Taylor expand the Einstein coupling constant $\chi(\rho)$ near this state. 
This will allow us to provide explicit expressions for all the quantities involved in the problem.
In the end, we want to stress that the present formulation is motivated by the idea that dark energy here does not correspond to a cosmological constant term, due to the quantum field vacuum in cosmology, but likely an evolutionary physical ingredient, as recently inferred by the DESI Collaboration \cite{desi}, see also \cite{Giare_Robust_DDE}.

Now, a natural expansion for the late Universe dependence of $\chi (\rho )$ can be taken as
\begin{equation}
	\chi (\rho ) = \chi_E \left( 
	1 - \frac{\rho^*}{\rho} + ...\right)\, ,
	\label{htx2x}
\end{equation}
where $\rho^*$ denotes a fiducial constant energy density. This formulation ensures a constant trend for $\chi$ when $\rho \gg \rho^*$, if this behavior were extrapolated at higher energy densities too.
{In fact, in what follows, we shall extrapolate the validity of the expression \eqref{htx2x} to the entire range of the Universe's energy density, motivated by its functional form: notably, it implies that the gravitational interaction progressively weakens as the energy density approaches that of the vacuum state, while it tends towards a constant value -- corresponding to the General Relativity limit -- for increasingly high energy densities.
Since this behavior aligns precisely with the phenomenological framework we are developing, it is a reasonable assumption to treat Eq. \eqref{htx2x} as a representative profile for $\chi (\rho )$ for the purpose of exploring the physical implications of our model.
}

Substituting the expression \eqref{htx2x} into the condition \eqref{htx1x} we get:
\begin{equation}
	\chi_E\left( 1 - \frac{\rho^*}{\rho_{\text{vac}}}\right) \rho_{\text{vac}} = 0 \rightarrow \rho_{\text{vac}}\simeq \rho^*\, .
	\label{htx3x}
\end{equation}
As a consequence, we see that the presence of a vacuum energy does not affect the Friedmannian dynamics of the Universe, since Eq. \eqref{htx11} now implies the basic relation:
\begin{equation}
	\sum _w\rho_w = \rho^* + 
	\sum_w \rho_w^s 
	\quad  , \, \rho_w^s \equiv \frac{\rho_w^0}{a^{3(1 + w)}} 
	\, , 
	\label{htx4x}
\end{equation}
where the label $s$ indicates the standard expression for the energy density with equation of state parameter $w$. 
We observe that, unlike in GR, a vacuum energy density here would never contribute to the Universe's expansion {but for an additional constant pressure term, see Sec. \ref{sec6}}.
Therefore, $\rho_{\text{vac}} \simeq \rho^*$ cannot be identified with the standard $\Lambda$CDM cosmological term, whose presence induces a distinct phenomenology.

\section{Implications for the late Universe dynamics}\label{sec5}

Let us now investigate the possible implication of our scenario when the late Universe dynamics is concerned.
While the fundamental evolution of the Universe, described by the Friedmann equations and particle trajectories of Section~\ref{sec3}, is not significantly altered by the varying Einstein coupling constant, we still observe an additional effect due to the vacuum energy density. Indeed, if the present-day Universe is close to the vacuum energy value, as assumed in the expression \eqref{htx2x}, for each matter component there is an associated extra constant pressure term given by $p_w^*=w\rho^*$, see Eq. \eqref{htx4x}. 

The energy density of the present-day Universe consists of three main contributions: matter ($w=0$), radiation ($w=1/3$), and dark energy ($w=-1$). 
The extra terms result in a net negative constant pressure 
\begin{equation}
	p^* = -\frac{2}{3}\rho^*\,,
	\label{htg1}
\end{equation}
which modifies the equation of state for dark energy. To capture this effect, we redefine the dark energy pressure as
\begin{equation}
	p_{de} = - \rho_{de} + p^* 
	\equiv w_{de}(\rho_{de})\rho_{de}\, , 
	\label{htg2}
\end{equation}
with a new effective parameter for dark energy
\begin{equation}
	w_{de} (\rho_{de}) \equiv - 1 - 
	\frac{2\rho^*}{3\rho_{de}}\, .
	\label{htg3}
\end{equation}
This interpretation is justified by the fact that we can think about dark energy as the extra component relevant today in addition to the dark matter. This leads to a modified cosmological dynamics with Friedmann equation
\begin{equation}
	H^2 = \frac{\chi_E}{3}\left( \rho_m + \rho_{de}\right)\, , 
	\label{htg4}
\end{equation}
where $\rho_m$ denotes the standard matter, and to continuity equations for the matter and dark energy components respectively: 
\begin{align}
	\frac{d\rho_m}{dz} &= \frac{3}{1+z}\rho_m\, , 
	\label{htg5}\\
	\frac{d\rho_{de}}{dz} &=\frac{3}{1+z}\left( 1 + w_{de}\right) \rho_{de}= 
	-\frac{2}{1+z}\rho^*\,.\label{htg6}
\end{align}
The dynamical expressions for the energy densities can be directly derived as solutions:
\begin{align}
	\rho_m(z) &= \rho_m^0(1+z)^3 \, , \\
	\rho_{de}(z) &= \rho_{\Lambda} - 2\rho^* \ln (1+z)\, , \label{htg8}
\end{align}
where $\rho_m^0$ is the present-day matter density and $\rho_{\Lambda}\equiv \rho_{de}(z=0)$ is the cosmological constant which corresponds to the standard $\Lambda$CDM model. Notably, Eq.~\eqref{htg8} exhibits a logarithmic correction, see also \cite{lemos-efstathiou}.

The Hubble parameter is also modified relative to the $\Lambda$CDM model. By introducing the standard normalization, its expression reads: 
\begin{equation}
	H^2(z)=H_0^2\left( \Omega_m^0(1+z)^3 + 1-\Omega_m^0 - \Omega^* \ln (1+z)\right) \, , 
	\label{htg9}
\end{equation}
where the density parameters for matter $\Omega_m^0$, cosmological constant $\Omega_{\Lambda}$, and vacuum energy $\Omega^*$ are defined as 
\begin{align}
	\Omega_m^0 &\equiv \frac{\chi_E\rho_m^0}{3H_0^2} \, ,\\
    \Omega_{\Lambda} &\equiv \frac{\chi_E \rho_{\Lambda}}{3H_0^2}= 1-\Omega_m^0\, ,\\
    \Omega^* &\equiv \frac{2\chi_E \rho^*}{3H_0^2}\,. 
	\label{htg11}
\end{align}
We recall that $H_0 \equiv H(z=0)$ is the Hubble parameter at redshift $z = 0$. 
Notably, the new parameter $\Omega^*$ is exactly twice the ratio of the vacuum energy density $\rho^*$ to the present-day critical density of the Universe. 
We proceed to confront this prediction with late Universe observables in the next Section.

\section{Model testing in the late Universe}\label{sec6}

We investigate here the possibility to constraint the three free parameters of the proposed model above, i.e. $H_0$, $\Omega_m^0$ and $\Omega^*$. In particular, we want to clarify if this latter parameter takes, from the data analysis, a value different from zero in one $\sigma$, which could validate the conjecture we discussed. 
For this purpose, we performed a parameter inference procedure on our model by using the publicly available sampler Cobaya \cite{cobaya} to implement Markov Chain Monte Carlo (MCMC) analysis. Specifically, we used a preliminary version of a code that will soon be publicly released [Giaré, Fazzari, in prep.].
We assess the convergence of our MCMC chains using the Gelman-Rubin $R-1$ parameter \cite{gelman_rubin}, and consider our chains converged when $R-1 < 0.01$. \\
To constrain $\Omega^*$, we conducted an MCMC analysis. We used a logarithmic prior over the range $[10^{-3}, 1]$, motivated by the expectation that $\Omega^*$ is positive due to its physical interpretation. We avoided sampling values up to null values because the model assumes an expansion regime where the fiducial value of $\Omega^*$ is expected to be of order unity. For model selection, we used the Bayesian factor $\ln B_{i,j}=\ln\mathcal{Z}{j}-\ln\mathcal{Z}{i}$, accounting for varying numbers of parameters. To interpret the significance of this factor, we adopted Jeffrey’s scale \cite{Bayes_trotta,jeffreys_scale}, which categorizes the evidence against the model as inconclusive if $0 < |\ln B_{i,j}| < 1.0$, weak if $1.0 < |\ln B_{i,j}|< 2.5$, moderate if $2.5 < |\ln B_{i,j}| < 5.0$, and strong if $|\ln B_{i,j}| > 5.0$.

\subsection{Datasets}
We used as observational datasets the main cosmological data at the background level, since we are investigating a late Universe modification of the $\Lambda$CDM model, that are:
\begin{itemize}
    \item \textit{Baryon Acoustic Oscillations (BAO)} -- BAO measurements consist of the transverse comoving distance ($D_M/r_d$), the Hubble horizon ($D_H/r_d$), and the angle-averaged distance ($D_V/r_d$), all normalized to the comoving sound horizon at the drag epoch $r_d$ \cite{BAO_cosmological,BAO_Melchiorri_1,BAO_Melchiorri_2,BAO_giare}. We use the DESI BAO measurements from the first-year data release, based on observations of the clustering of the Bright Galaxy Sample (BGS), the Luminous Red Galaxy Sample (LRG), the Emission Line Galaxy (ELG) Sample and the combined LRG+ELG sample, quasars, and the Lyman-$\alpha$ forest as summarized in Table I of Ref.~\cite{desi}. The data span the redshift range $0.1 < z < 4.16$, and we account for the correlation between measurements of $D_M/r_d$ and $D_H/r_d$. The sound horizon is calibrated using Planck data, assuming a Gaussian prior of $r_d=(147.09\pm0.26)$ Mpc, as reported in Table 2 of Ref.~\cite{Planck2018}. We refer to this dataset as "DESI". 
    \item \textit{Cosmic Chronometers} -- measurements of the expansion rate $H(z)$ from so-called cosmic chronometers (CC), i.e. the differential ages of massive, early-time, passively-evolving galaxies \cite{Jimenez_247,CC_borghi}. For our analysis, we use 15 data points reported in Refs. \cite{Moresco_4,moresco_248,Moresco_249} in the range $0.1791 < z < 1.965$. While more than 30 CC measurements are technically available, we focus our analysis on a subset where full estimates of the covariance matrix's non-diagonal terms and systematic contributions, as outlined in Refs. \cite{Moresco_250,Moresco_251}, are accessible. Additionally, we exclude some earlier measurements due to concerns expressed in Ref. \cite{Ahlstr_m_Kjerrgren_252}, which do not apply to our selected data. We note that including the other CC measurements is unlikely to significantly impact our results, as our chosen sample already includes some of the most precise and reliable measurements. We refer to this set of 15 measurements as "CC", and the corresponding data is publicly available in the repository \cite{CC_git}.
    \item \textit{Type Ia Supernovae (SNe Ia)} -- distance moduli measurements \cite{riess2016,scolnic2018} used in two different compilations: 
    \begin{itemize}
    \item[-] PantheonPlus sample \cite{Scolnic_2022,Brout_2022} that consist of 1701 light curves for 1550 uncalibrated SNe Ia spanning a redshift range of $0.01$ to $2.26$. This dataset is referred to as "SN". 
    \item[-] PantheonPlus with the SH0ES Cepheid host distances used to calibrate the SN Ia sample \cite{SH0ES}. We denote the SH0ES calibrated sample as "SH0ES". 
    \end{itemize}
    For both cases the likelihood has been taken from the public repository \cite{SN_git}.
\end{itemize}

\subsection{Results}
Here we discuss the results for the parameter inference procedure. In Table \ref{Tab:Results_Combined}, we present the
datasets used and the inferred parameter values for both the $\Lambda$CDM model and our proposal. 

\begin{figure*}[h]
    \centering    
    \includegraphics[width=0.7\textwidth]{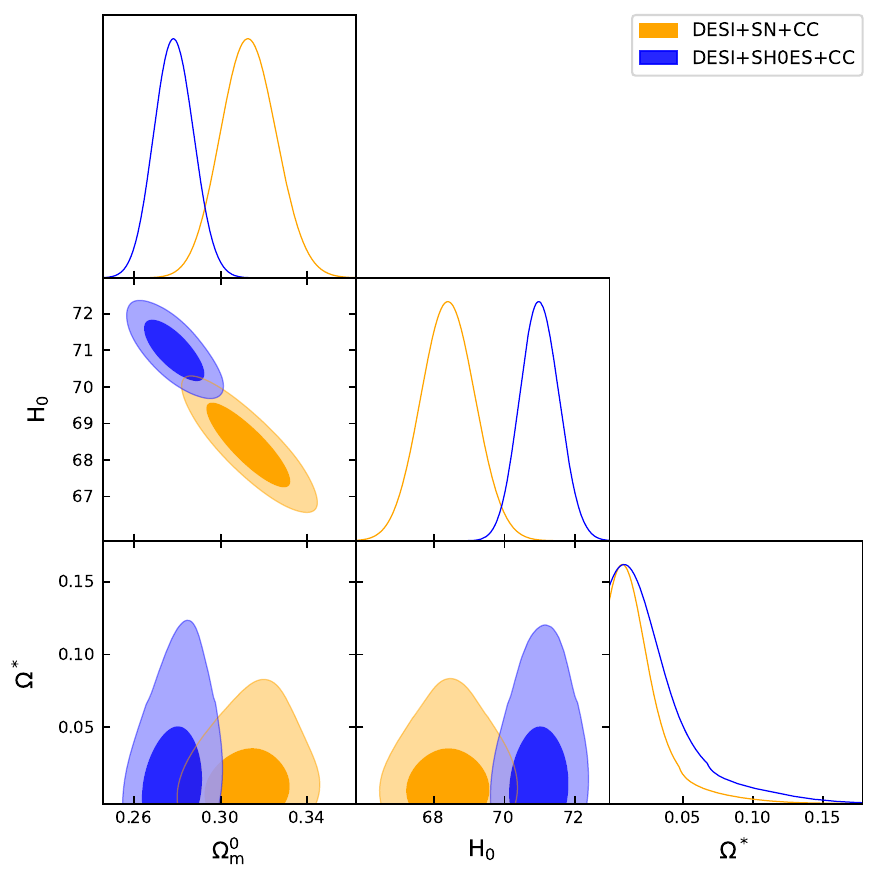}
    \caption{One-dimensional posterior probability distributions and two-dimensional $68 \%$ and $95\%$ CL contours for the free parameters of the model $H_0$, $\Omega_{m}^0$, $\Omega^*$, as inferred by the two dataset combination listed in the legend.
    \label{fig:contour_confronto_1}}
\end{figure*}

We show in Figure \ref{fig:contour_confronto_1} the one-dimensional posterior probability distributions and two-dimensional $68\%$ and $95\%$ CL contours for the free parameters of the model.
From the results, we observe that for both dataset combinations, the mean value of $\Omega^*$ is compatible with zero within $1\sigma$. Notably, the best-fit value for $\Omega^*$ is positive and the peak of its posterior distribution is significantly different from the lower-limit of the prior range. The two datasets are consistent in their best-fit value of $\Omega^*$. Moreover, the statistical analysis indicates that our model is indistinguishable from the $\Lambda \mathrm{CDM}$ model in terms of their fit to the data.

\noindent We observe that the uncertainties associated with $\Omega^*$ are relatively large, highlighting the challenges in tightly constraining this parameter.

\begin{table*}[h]
\centering
\resizebox{\textwidth}{!}{
    \begin{tabular}{lcccccc}
     \hline\noalign{\smallskip}
     Model & Dataset & $\mathbf{H_0}$[km s$^{-1}$Mpc$^{-1}$] & $\mathbf{\Omega_m^0}$ & $\mathbf{\Omega^*}$ & $\Delta \chi ^2$ & $\Delta \ln B_{i, \Lambda \mathrm{CDM}}$ \\
    \noalign{\smallskip}\hline\noalign{\smallskip}
    $\Lambda$CDM & DESI+SN+CC    & $68.36\pm 0.69$   & $0.312\pm 0.012$   & $-$       &   $-$    &   $-$  \\
    & DESI+SH0ES+CC & $70.30\pm 0.58$   & $0.285\pm 0.009$ & $-$    &   $-$    &   $-$              \\
   \noalign{\smallskip}\hline\noalign{\smallskip}
    Model
      & DESI+SN+CC    & $68.41\pm 0.72$   & $0.313\pm 0.012$   & $0.015^{+0.011}_{-0.025}$  &   $-0.003$    &   $0.036$  \vspace*{0.2em}\\
    
      & DESI+SH0ES+CC & $71.00\pm 0.52$   & $0.278\pm 0.009$ & $0.022^{+0.012}_{-0.037}$   &   $-0.042$    &   $-0.197$\\
  \noalign{\smallskip}\hline
    \end{tabular}
}
\caption{Mean values and associated uncertainties for the inferred parameters from the MCMC analysis for the $\Lambda$CDM and our model. We report the quantity $\Delta \chi^2 = \chi^2_{i} - \chi^2_{\Lambda \mathrm{CDM}}$ as well as the difference between the Bayes factors of $\Lambda \mathrm{CDM}$ and our model, defined as $\Delta \ln B_{i, \Lambda \mathrm{CDM}} = \ln B_i - \ln B_{\Lambda \mathrm{CDM}}$.} 
\label{Tab:Results_Combined}
\end{table*}

\begin{figure*}[h]
    \centering
    \includegraphics[width=0.62\textwidth]{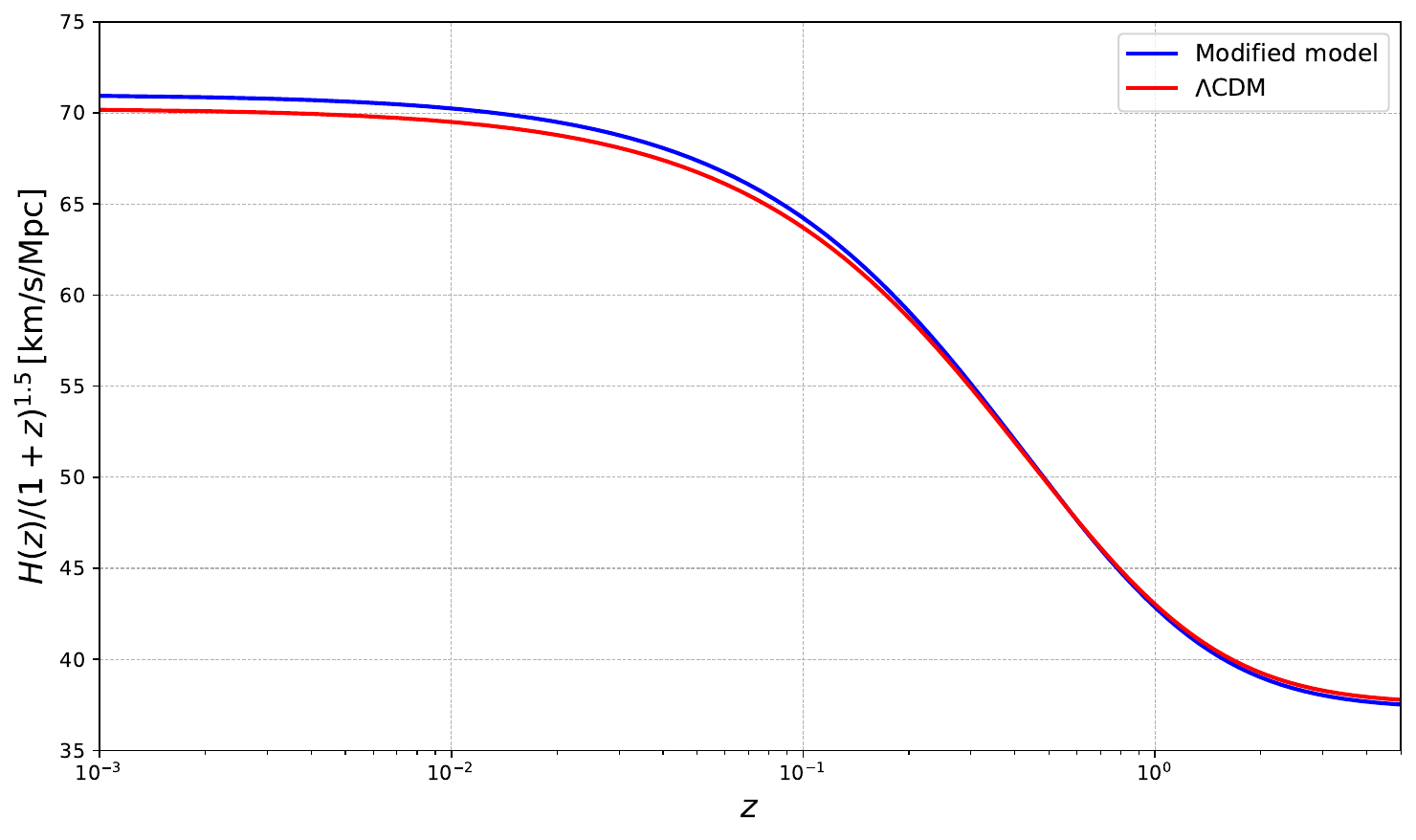}
    \caption{Evolution of the normalized Hubble parameter $H(z)/(1+z)^{1.5}$ for the modified model (blue line) and the $\Lambda$CDM model (red line), using the best-fit parameter values in Eq. \eqref{eq:best-fit} obtained from the DESI+SH0ES+CC dataset combination }.
    \label{fig:H_z}
\end{figure*}

Figure \ref{fig:H_z} shows the Hubble parameter profile for the $\Lambda$CDM model and our modified scenario using the best-fit values of the parameters obtained from the dataset combination DESI+SH0ES+CC:

\begin{equation}
    \begin{aligned}
    H_0^{\Lambda\mathrm{CDM}}&= 70.25 , \; \; \Omega_m^{0,\,\Lambda\mathrm{CDM}}= 0.286 \\
    H_0&= 71.03, \; \; \Omega_m^{0}= 0.276 ,\;\; \Omega^*= 0.004 \,.
    \label{eq:best-fit}
    \end{aligned}
\end{equation}

\noindent We observe that our model provides a slightly higher value of $H_0$ compared to the $\Lambda$CDM one, and that the two curves overlap at $ z \sim1$. 
 Focusing on the values of $H_0$ and $\Omega_m^0$, we observe that adopting the SH0ES calibration results in higher $H_0$ values, accompanied by a lower $\Omega_m^0$. This effect is also reported in \cite{Scolnic_2022}. Furthermore, the combinations for $\Omega_m^0 h^2$ (where $h=\frac{H_0}{100\; \text{km} \; \text{s}^{-1}\text{Mpc}^{-1}}$) are consistent with each other within $1\sigma$ and agree with the Planck value reported in \cite{Planck2018}.
Additionally, the posteriors on $r_d$ in both cases align well with the Planck constraint provided as Gaussian prior to calibrate the sound horizon for BAO measurements.

\section{Concluding remarks}\label{sec7}

We have proposed a reformulation of GR applied to the Universe dynamics based on a running of the gravitational coupling constant with the density. 
We examined the consequences of the Bianchi identities on the conservation law for the matter content of the Universe, as a superposition of perfect fluids in the standard framework. 
In this revised scenario, both the Friedmann equation and the free particle motion retain formally the same expression with respect to the standard model, but an additional constant pressure term arises.

We constructed an explicit expression for $\chi(\rho)$ by means of a Taylor expansion for $\rho \simeq \rho_{vac}$, which was then extended to the whole range of density so providing a viable paradigmatic assignment for the proposed theory.

In this regime, the model provided both an explicit expression for the Einstein coupling constant as a function of the Universe's energy density and the corresponding energy density for the superposition of perfect fluids. 
In view of the vacuum energy density problem, this result implied an additional term to the total energy density (which, in the absence of a standard cosmological constant, represents its asymptotic future value), while it factors out of the Friedmann equation and free particle motion.

This formulation also introduced an additional constant pressure term in the dynamics; this ``anomalous'' contribution can be interpreted as a modified equation of state for dark energy, resembling a Chaplygin-like gas behavior \cite{chaplyging_1,chaplyging_2}. 
Such effect introduced a negative logarithmic  correction to the Hubble parameter, whose magnitude is equal to twice the ratio of the vacuum energy density to the critical density of the present-day Universe, see Eq.~\eqref{htg11}.

The revised late Universe dynamics was then tested against observational data from sources detected at redshifts less than a few units, i.e. SNe Ia, CC and BAO distances. 
By constraining the coefficient of the logarithmic term to be positive (implying a positive vacuum energy density), a Bayesian inference procedure using an MCMC method revealed that its value is compatible with zero within $1 \, \sigma$.
In other words, no significant evidence emerges of a nonzero value of $\Omega^*$. 

Based on the theoretical and numerical analysis conducted, we can claim that either the vacuum energy density is exactly zero within our present-day data sensitivity (so that GR is fully recovered), or that the present status of datasets prevents a significant constraining of its value in the modified scenario.

We conclude by observing that the contribution of the logarithmic term to a possible attenuation of the 
Hubble tension \cite{montani_hubbletension_1,montani_hubbletension_2,montani_hubbletension_3,montani_hubbletension_4,montani_hubbletension_5,montani_hubbletension_6,dainotti_montani_divalentino,divalentino_hubble_tension_2,giare_Hubble_tension_1} is rather weak. 
Only in one case (i.e. considering DESI+ SH0ES+CC), the model analyzed reduces the tension with respect to the result in \cite{SH0ES} at $1.75\, \sigma$.
This alleviates the tension with respect to the $\Lambda$CDM model in the same dataset combination of $0.55\, \sigma$.    

While we recognize that the model does not outperform $\Lambda$CDM in terms of statistical
preference, we believe that it provides an illustrative example of how vacuum energy could
be dynamically deactivated at late times, with minimal modifications to the Friedmann
equations. 
Further analyses and possibly higher data resolution are needed to draw insights on the reliability of such proposal.

\paragraph{\textbf{Data Availability Statement}}
My manuscript has associated data in a
data repository. [Author’s comment: The data supporting the findings of this study are publicly available at the repositories \cite{CC_git,SN_git,William_git}.]

\paragraph{\textbf{Code Availability Statement}}
Code/software will be made available on
reasonable request. [Author’s comment: The code generated and analyzed during the current study is available from the authors on reasonable request.]

\begin{acknowledgement}
\noindent We would like to thank very much E. Di Valentino and W. Giarè for the excellent advice on the model testing setup. In particular, E.F. is grateful to W. Giarè for his technical support for code running and for his public GitHub repository \cite{William_git} used for statistical analysis. 
G. Maniccia would like to thank the Institute for Quantum Gravity, Friedrich-Alexander-Universit\"at Erlangen-N\"urnberg for hospitality and Sapienza University of Rome for funding.
E. F. acknowledges the IT Services at The University of Sheffield for providing High Performance Computing resources. \\
This article is based upon work from COST Action CA21136 – “Addressing observational tensions in cosmology with systematics and fundamental physics (CosmoVerse)”, supported by COST (European Cooperation in Science and Technology). Authors E.F. and A.M. are supported by "Theoretical Astroparticle Physics" (TAsP), iniziativa specifica INFN. 
A.M.  and E.F. acknowledge support by the research grant number 2022E2J4RK "PANTHEON: Perspectives in Astroparticle and Neutrino THEory with Old and New messengers" under the program PRIN 2022 funded by the Italian Ministero dell’Universit\`a e della Ricerca (MUR) and by the European Union – Next Generation EU.

\end{acknowledgement}

\bibliographystyle{spphys} 
\bibliography{htgbiblio.bib}

\end{document}